# Table of Contents Graphic

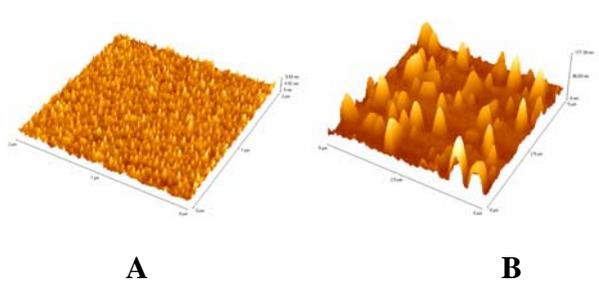

**A**                  **B**

50 nm $Nd_{0.5}Sr_{0.5}CoO_3$ thin film "as deposited" (A) and "annealed" (B)

# Brief Summary

Single phase thin films of $Nd_{1-x}Sr_xCoO_3$ ($x$=0, 0.2 and 0.5) perovskite with nanocrystalline morphology have been deposited on single crystalline substrates ($SrTiO_3$ and $LaAlO_3$) by means of rf-magnetron sputtering. Influence of the substrate nature, thickness and thermal treatments have been studied.

# RF Sputter Deposition of Epitaxial Nanocrystalline $Nd_{1-x}Sr_xCoO_3$ Thin Films

*Lorenzo Malavasi[1,*], Eliana Quartarone[1], Carla Sanna[2], Nathascia Lampis[2], Alessandra Geddo Lehmann[2], Cristina Tealdi[1], Maria Cristina Mozzati[3], and Giorgio Flor[1]*

[1]Dipartimento di Chimica Fisica "M. Rolla" and INSTM, Università di Pavia, Viale Taramelli 16, 27100 Pavia, Italy.

[2]Dipartimento di Fisica – Università di Cagliari - Cittadella Universitaria St. Pr.le Monserrato-Sestu km. 0.700, I-09042 Monserrato (Ca), Italy.

[3]CNISM, Unità di Pavia and Dipartimento di Fisica "A. Volta", Università di Pavia, Via Bassi 6, I-27100, Pavia, Italy.

*Corresponding Author: Dr. Lorenzo Malavasi, Dipartimento di Chimica Fisica "M. Rolla", INSTM, Università di Pavia, V.le Taramelli 16, I-27100, Pavia, Italy. Tel: +39-(0)382-987921 - Fax: +39-(0)382-987575 - E-mail: lorenzo.malavasi@unipv.it

# ABSTRACT

In this paper we report the deposition of epitaxial thin films of $Nd_{1-x}Sr_xCoO_3$ with $x$=0, 0.2 and 0.5 on single crystalline substrates ($SrTiO_3$ and $LaAlO_3$) carried out by means of rf-magnetron sputtering. The deposited films are all completely oriented and epitaxial and characterized by a nanocrystalline morphology. As-deposited films have an average roughness around 1 nm while after the thermal treatment this increases up to 20 nm while preserving the nanocrystalline morphology. All the films deposited on $SrTiO_3$ have shown to be under a certain degree of tensile strain while those on the $LaAlO_3$ experience a compressive strain thus suggesting that at about 50 nm the films are not fully relaxed, even after the thermal treatment. For the $x$=0.2 composition three different thickness have been investigated revealing an increased strain for the thinner films.

KEYWORDS: cobaltite, perovskite, thin films, rf-sputtering, x-ray diffraction.

# Introduction

Cobalt-containing perovskite-type oxides, particularly La-rich oxides, have been the subject of intense research mainly due to the possibility of optimizing their structural and physical properties by doping. The range of possible application for these oxides is wide, extending from components in solid oxides fuel cells (SOFCs), oxygen separation membranes and electrochemical reactors, to sensor devices based on their ability to catalytically oxidize CO and $CH_4$ and reduce NO.[1-6]

In addition, perovskite cobaltites have received further attention after the observation of elevated Seebeck coefficients in related layered systems such as $GdBaCo_2O_{5+\delta}$[7-9] which suggest their possible application as thermoelectric materials (TE). Recent reports showed that doped perovskite cobaltites have relatively high figures of merit ($Z$) and this directly correlates to the different spin states available for the cobalt ions[10-13]. As a consequence, the TE materials research is the latest field where cobaltites appeared as promising and new compounds.

Their magnetic and magnetoresistive properties are also of recent interest. In particular, a debate has been opened regarding the actual spin-state of Co in lanthanum cobaltites. In fact, it has been shown that in the range from 5 to 1000 K the Co ions pass through three different spin-states (low-spin LS, intermediate-spin IS, and high-spin HS) which are intimately connected to the structural as well as internal parameters such as the metal-oxygen bond lengths[14]. The spin state, in turn, affects the physical properties such as transport, magnetic and optical properties[15-18] acting as a sort of "Jahn-Teller switch", where for certain spin configurations the J-T distortion, for the same Co valence state, is suppressed. A general gradient approximation (GGA) study due to Knižek[19] demonstrated that the relative stability of IS and LS depends on the Co-O distances and angles with longer bonds and more open Co-O-Co angle favouring the IS state. It is clear that the internal parameters, such as bond lengths and angles, play a major role in defining the cobaltites physical properties.

Among RECoO$_3$ (RE=rare earth) compounds, LaCoO$_3$ and NdCoO$_3$ have been object of previous investigation both as pure compounds and considering the role of divalent dopant (Sr) concentration[20-23]. Much of the previous work focussed on bulk or single crystalline materials and less on thin films preparation and characterization which, however, is the useful "physical form" for sensing and catalytic as well as TE micro-devices applications. Moreover, it would be desirable to make available nanocrystalline thin films where the high surface-to-volume ration can enhance the cobaltites physical properties.

The current literature concerning with the synthesis of cobalt containing perovskite thin films is not rich and mainly devoted to the LaCoO$_3$ and Sr-doped LaCoO$_3$ whose preparation has been carried out by means of sol-gel[24-26], screen-printing[27], pulsed laser deposition (PLD)[28], spray pyrolysis[29] and hybrid CVD/sol-gel route[30]. Thin films of other phases with a different lanthanide on the A-site, such as the object of this work, have not been considered in the previous literature. However, other materials, such as the NdCoO$_3$ perovskite, have already shown to be highly interesting materials for both magnetic[18,20], sensing[31-33], catalytic[34,35], SOFC[36,37] and thermoelectric properties[38].

In this paper we report the synthesis and structural and morphological investigation of oriented epitaxial thin films of Nd$_{1-x}$Sr$_x$CoO$_3$ with $x$=0, 0.2 and 0.5 on single crystalline substrates (SrTiO$_3$ and LaAlO$_3$). The Sr-dopant has been chosen since it has shown to be one of the most soluble[39] and one of the most favourable dopants in terms of structure distortion, with a tolerance factor close to 1. This leads to low-distorted phases where the Co-O-Co hole hopping is favoured. Increasing Sr-concentration gives origin to higher Co valence state and oxygen vacancies concentration, which are thought to be the active sites for gas adsorption. We stress that among the current literature, to the best of our knowledge, this is the first work reporting the deposition of a cobalt containing perovskites thin films by means of rf-sputtering and, in addition, the first paper reporting the preparation of oriented Nd$_{1-x}$Sr$_x$CoO$_3$ thin films.

# Experimental Section

Powder samples of $NdCoO_3$, $Nd_{0.8}Sr_{0.2}CoO_3$ and $Nd_{0.5}Sr_{0.5}CoO_3$ have been prepared by conventional solid state reaction from the proper stoichiometric amount of $Nd_2O_3$, $Co_3O_4$, and $SrCO_3$ (all Aldrich ≥99,9%) by repeated grinding and firing for 24 h at 900-1050 °C.

Thin films were deposited onto single-crystalline $SrTiO_3$ (001), and $LaAlO_3$ (100) (Mateck) by means of off-axis rf-magnetron sputtering (Rial Vacuum). The gas composition in the sputtering chamber was argon and oxygen (16:1) with a total pressure of $4\times10^{-6}$ bar. The substrate was heated at 700°C and rotated during deposition. The rf-power was set to 150 W. After the deposition the films were annealed at 900°C in pure oxygen for 30 minutes. The chemical composition of starting powders and thin films was checked by means of electron microprobe analysis (EMPA) which confirmed their correct cation ratio.

XRR data have been collected by using a Bruker D8 advance reflectometer equipped with a Göbel mirror. The Cu K$\alpha$ line of a conventional X-ray source powered at 40 kV and 40mA was used for the experiment. The grazing incidence specular reflectivity curves were recorded with a $\theta$-$\theta$ scan in the 0–3° range. X-ray diffraction (XRD) patterns of starting powders were acquired on a Bruker D8 Discover diffractometer equipped with a Cu anode in a $\theta$-$2\theta$ geometry. High resolution reciprocal space maps around symmetric and asymmetric reflections and ω-scans (rocking curves) on the epitaxial layers were performed on a 4-circles "Bruker D8 Discover" diffractometer (Cu anode) equipped with Goebel mirror for parallel beam geometry and pure K$\alpha$ selection and with a 2-bounces (V-Groove) monochromator (Ge 022) for K$\alpha_2$ elimination.

AFM images (256×256 pixels) were obtained with an AutoProbe CP microscope (ThermoMicroscopes-VEECO), operating in contact mode (C-AFM), by means of sharpened silicon tips onto V-shaped cantilevers (resonance frequency: 15 kHz; force constant: 0.03 N/m). For each analysed film, scans of 5 μm × 5 μm and 2.0 μm × 2.0 μm have been carried out with a scan rate ranging from 1.0 to 1.5 Hz. A standard 2$^{nd}$ order flatten processing of the images has been performed in order to correct the scanner non-linearity.

## Results and Discussion

Figure 1 reports the refined X-ray diffraction pattern of the $Nd_{0.8}Sr_{0.2}CoO_3$ target material, chosen as a representative example. For all the three samples, namely $NdCoO_3$, $Nd_{0.8}Sr_{0.2}CoO_3$ and $Nd_{0.5}Sr_{0.5}CoO_3$, the X-ray diffraction patterns can be perfectly refined considering an orthorhombic unit cell (space group n. 62, *Pnma*) with lattice constants reported in Table 1.

Doping the Nd-site with of Sr induces a progressive expansion of the unit cell from 215.59(1) Å$^3$ to 220.67(2) Å$^3$. This relatively small expansion with respect to the difference in the ionic radii which, for the same coordination (12), are 1.27 Å for $Nd^{3+}$ and 1.44 Å for $Sr^{2+}$, is due to the concomitant oxidation of Co ions as the Sr-doping occurs. As a matter of fact, we expect that the Sr-doping will increase the hole concentration according to the following equilibria, which take into account the cation replacement and the (partial or total) compensation of oxygen vacancies with external oxygen:

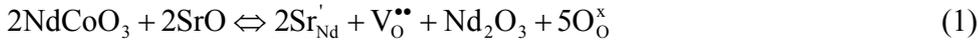

$$2NdCoO_3 + 2SrO \Leftrightarrow 2Sr'_{Nd} + V_O^{\bullet\bullet} + Nd_2O_3 + 5O_O^x \quad (1)$$

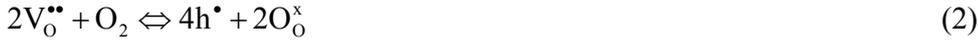

$$2V_O^{\bullet\bullet} + O_2 \Leftrightarrow 4h^\bullet + 2O_O^x \quad (2)$$

Starting from the pure and Sr-doped $NdCoO_3$ targets we deposited a series of thin films on single-crystalline substrates. We chose to grow the films on cubic $SrTiO_3$ (001) and $LaAlO_3$ (001). The choice has been done considering the lattice constants of the target materials with respect to the parameters of the substrate. In particular, the cubic axis for $SrTiO_3$ is 3.905 Å while for $LaAlO_3$ the pseudo-cubic axis is 3.785 Å. Considering a pseudo-cubic cell for the Sr-doped neodymium cobalt perovskites, we may expect an *average* parameter around 3.78 Å for the $NdCoO_3$, ~3.79 Å for the $Nd_{0.8}Sr_{0.2}CoO_3$ and ~3.81 Å for the $Nd_{0.5}Sr_{0.5}CoO_3$, respectively. So, in principle, we should be able to look at the role of substrate nature on the growth of the films and on the physical properties induced by possible modulation of the lattice parameters (*i.e.* strain).

The deposited films have been, first of all, characterized through X-ray reflectivity (XRR) to determine their thickness. The estimated standard deviation in the determination of the film thickness by means of XRR is around 4-5%. Figure 2 reports, as an example, the XRR spectrum for the pure $NdCoO_3$ thin film deposited on STO for about 30 minutes. In the inset it is highlighted a small part of the spectrum in order to put in evidence the Kiessig fringes which originate from the interference between successive layers of the film. The separation between two successive maxima in the curve is a direct and reliable measurement of the thin film thickness, $t$. However, the $t$ value has been calculated from a fit of the experimental curve to a model curve by means of the LEPTOS software (Bruxer AXS). It turned out that the film thickness is around 50 nm, thus indicating a deposition rate of ~1.7 nm/min. For the three cobaltites, namely $x$=0, 0.2 and 0.5, we deposited 50 nm thin films on STO and LAO substrates. For the $x$=0.2 composition we also deposited thinner films, i.e. ~8.5 and 17 nm, in order to look at the role of film thickness. For all the deposited films the roughness estimated from the XRR are around or lower than 1 nm prior to oxygen annealing. After the thermal treatment the film roughness is significantly enhanced and reach, usually, values of few nm for the 50 nm thin films. Some more details about the samples roughness is given later in the text when discussing the AFM data. We note that in this work we were mainly interested in looking at thin films (< 100 nm), where the substrate-induced effects are more significant but the preparation route detailed here can be easily applied to the deposition of thicker films of hundreds of nanometers.

A full x-ray diffraction characterization has been carried out on the as-deposited as well as annealed thin films of all the three compositions. In the following we will show some representative examples of the general behaviour shown by the samples.

Let us look first at the role of post-deposition annealing treatment. Figure 3 reports the X-ray diffraction patterns for the $Nd_{0.8}Sr_{0.2}CoO_3$ 50 nm film (labelled as NSCO in Figure 3) measured immediately after the deposition (black line), and after the oxygen annealing (blue line), respectively; in addition, vertical red bars represent the STO peaks (also labelled in the Figure) while the green lines are relative to the orthorhombic $Nd_{0.8}Sr_{0.2}CoO_3$ peaks.

First of all we note that the thin films are highly oriented since in their X-ray patterns are clearly visible the only peaks related to those of the substrate. Before the annealing treatment the full-width-at-half-maximum (FWHM) of the peak centred around 48° is 0.352°; after the thermal treatment the peak width slightly reduces to 0.327°, thus suggesting the crystallinity of the films is not significantly improved by this relatively short treatment. From the X-ray patterns of Figure 3 we calculated the out-of-plane parameter for the 50 nm $Nd_{0.8}Sr_{0.2}CoO_3$ thin film before and after the annealing treatment. In the first case the out-of-plane coordinate is 3.781(3) Å while after the heat treatment the parameter slightly contracts to 3.771(3) Å. This is most probably due to the partial oxidation of the cobalt ions with the creation of smaller oxidised species such as $Co^{4+}$. This effect has been found to be a general trend for all the films.

The role of the thermal treatment on thin film morphology was studied. Figure 4 shows the comparison between the morphology of the 50 nm thin film of $Nd_{0.8}Sr_{0.2}CoO_3$ after the deposition ("as-deposited" sample) and after the thermal treatment ("annealed" sample). The as-deposited sample has a very low roughness of about 0.8 nm and is formed by nanocrystalline grains with an average size of about 10-15 nm characterized by a relatively narrow grain size distribution. The treatment at 900°C leads to a significant increase of both the grain size and the roughness, which is now around 5 nm. Peak in the grain size distribution after the thermal treatment is around 80 nm. Let us note, however, that the grains of the annealed sample are actually made of smaller grains of tents of nm; the high temperature treatment led to the formation of bigger island as a consequence of the coalescence of the previous, smaller, islands.

Concerning the thin films orientation we can not be conclusive based on these XRD data. In fact, considering the pseudo-cubic cell parameters derived from the orthorhombic ones and assuming that the thin films adopt the *Pnma* crystal structure of the target material, two orientations of the films, with respect to the substrate, can be found: i) [010]-orientation, and thus the peaks in Figure 3 correspond to the (020) and (040) reflections of the cobaltite, and ii) [101]-orientation, with peaks in the pattern corresponding to the (101) and (202) cobaltite planes. This point is illustrated in more details in Figure 1-SI available as Supporting Information, where, for a wider specular scan, the two sets of Miller indices for the two different epitaxial orientations of the $Nd_{0.8}Sr_{0.2}CoO_3$ film are indicated. Finally, we may not exclude that the substrate induced the growth of more symmetric films with respect to the target materials. In particular, cubic or more probable, tetragonal symmetry may not be ruled out based on these data. In the following we will use a pseudocubic approximation to evaluate the relaxation degree of the films.

Let us now pass to discuss the role of Sr-doping on the thin films structure and morphology. Figure 5 shows the X-ray diffraction patterns for the three 50 nm thin films grown on STO (001) substrate for those two regions of the patterns where the (002) and (004) cubic diffraction peaks due to the perovskite are present. The out-of-plane lattice parameter for the three films is reported in Table 2. As can be seen, by increasing the Sr-concentration the parameter tends to expand, in accordance with the higher ionic radius of Sr with respect to Nd. We may note that the out of plane parameter for the three films is smaller than the pseudo-cubic parameter calculated from the bulk lattice constants. This is most probably directly connected to the tensile strain induced by the substrate, that is, a decreasing in the growth direction and expanding in the plane, which has an in-plane parameter of 3.90 Å.

To evaluate the in plane lattice parameters and the relaxation degree of the films $R = \frac{a_{strained} - a_{substrate}}{a_{bulk} - a_{substrate}}$ - in pseudocubic (tetragonal) approximation - high resolution reciprocal space maps (RSM) have been performed around the asymmetric cubic reflection $(103)^+$ (grazing exit). The results for the 50 nm thick Sr-doped films with $x=0, 0.2, 0.5$ are shown in Figure 6A-C. The estimated in-plane lattice parameters are $a=3.847(6)$ Å, $3.836(4)$ Å and $3.820(3)$ Å, leading to a relaxation degree $R$ of about 50%, 60% and 90% respectively for the three compositions. We note that even if the film of composition $Nd_{0.5}Sr_{0.5}CoO_3$ appears to be to most relaxed among the analyzed samples, its epitaxial quality is still high as shown by the narrow rocking curve on the (002) pseudocubic reflection (Figure 7a) and also by the RSM around the symmetric (002) cubic reflection (Figure 7b) which shows that the width of the Bragg diffusion along the L direction (which depends on the out of plane texture) is the same for the film and of the oriented single crystalline substrate. We also note in the (002) map the absence of the film truncation rod intensity, which indicates a three dimensional growth in agreement with AFM results.

The same structural investigation has been carried out for the three 50 nm films on LAO (001). Figure 8 reports the results for the undoped sample, i.e. $NdCoO_3$. First we note that also with the LAO substrate the film growth is completely oriented. In this case the lattice parameters of the substrate and those of the cobalt perovskites are closer with respect to the STO substrate. The pattern in Figure 8 shows the presence of just two peaks exactly located at the position of the (001) and (002) reflections of the LAO substrate. In the inset of the Figure it is presented the comparison of the diffraction patterns of the $NdCoO_3$ film on LAO (red line) and of the substrate (black line). As can be appreciated, the only difference between the two patterns is the lack, in the film pattern, of the clear $K\alpha_1/K\alpha_2$ separation which is clearly visible in the substrate pattern. This means that the film has grown with lattice constants so close to those of the LAO substrates that is not possible to discriminate between them. This is also due to the FWHM of the films being of the order of 0.3° with respect to the FWHM of the single crystal peaks whish is around 0.05°. Overall, this result means that the out-of-plane parameter for the $NdCoO_3$ film on LAO is 3.785(5) Å and considering that the in-plane parameter for the LAO material is 3.785 Å this suggests that the perfect match of the lattice parameters between the film and substrate leads to the growth of a nearly cubic film.

For the other two compositions (patterns not shown), i.e. $x$=0.2 and 0.5, the out-of-plane parameter is bigger. For $x$=0.2 it is about 3.816(3) Å, to be compared to the pseudo-cubic parameter of the bulk phase being ~3.79 Å. In this case the film grows under the effect of a compressive strain, that is, an expanding in the growth direction and a decreasing in the plane. For the $x$=0.5 sample the effect is even higher, with a out-of-plane parameter close to 3.820(2) Å.

We also note that different Sr-dopings lead to different morphologies. Figure 9 shows, as an example, the AFM measurements carried out on the $x$=0.5 50 nm film deposited on STO (001). This has to be compared to Figure 4, which reports the same data for the $x$=0.2 composition. As can be seen, by increasing the Sr-doping, the surface roughness increases from 0.8 nm ($x$=0.2) to 1.1 nm ($x$=0.5) and it is clearly composed by very tiny grains which are practically undetectable on the surface of the $x$=0.2 film. A greater difference between the two samples is found looking at the surfaces of the annealed samples. The difference in the surface roughness between the two compositions is now higher: ~4.6 nm for the $x$=0.2 film and ~23 nm for the $x$=0.5 film. Also the morphology is significantly different, with elongated island for lower Sr-doping with respect to the round-shaped found for $x$=0.5. This difference is interesting and is most probably correlated to the different growth properties (both adsorption and diffusion processes) of the two compositions induced by the presence of a higher doping level of Sr. More details about this topic are beyond the scope of the present paper and will be considered in the future. However, for practical applications in which surface properties are crucial, the knowledge of this doping-dependence of both roughness and morphology has to be taken into account.

For the $Nd_{0.8}Sr_{0.2}CoO_3$ sample grown on STO we tried to look at the role of film thickness on the structural properties. Figure 10 shows the X-ray diffraction pattern for three thicknesses: 8.5, 17 and 50 nm. The out-of-plane parameter for the thicker film is, as mentioned above, 3.771(3) Å, while for the other two films it is slightly reduced to about 3.753(5) Å. For both 8.5 and 17 nm the diffraction peaks looks also wider with respect to the 50 nm sample being the 8.5 nm the one with the highest FWHM (ca. 1.2°). The presence of a smaller lattice parameter for the 8.5 and 17 nm samples suggest a stronger effect of the substrate on the film which induces a higher degree of strain; in both cases the samples are less relaxed with respect to the 50 nm sample. The presence of broader diffraction peaks for the thinner samples may be the result of the presence of a relaxation gradient or/and of an interface substrate/film layer contribution where the chemical composition of this layer is different with respect to that of the film "bulk".

Finally, the influence of substrate temperature during thin film deposition is put in prominence through Figure 11. Here, it is reported the XRD pattern of a $NdCoO_3$ thin film deposited while heating the substrate to 400°C instead of heating to 700°C, as done for the other films considered in this work; all the other deposition parameters were kept constant. As can be appreciated, beside the cobaltite phase other intense peaks appear in the pattern (marked with an asterisk). At present we are not able to undoubtedly associate these peaks to a precise phase. The peak located at about 38° might be the (102) and (201) reflections of the orthorhombic cobaltite, thus indicating that in order to obtain full oriented films the deposition has to occur by heating the substrate to high temperatures. Anyway, peak at around 65.5° can not be related to the cobaltite structure. As a consequence, another possibility is that these peaks originate from a second phase of still uncertain composition (which is not, however, any of the simple metal oxides of Co or Nd). We also remark that the FWHM of the $NdCoO_3$ peaks is around 0.55° which strongly indicates that higher deposition temperatures are effective in improving the thin film crystallinity.

# Conclusion

In this paper we report, to the best of our knowledge, the first deposition of epitaxial thin films of $Nd_{1-x}Sr_xCoO_3$ with $x$=0, 0.2 and 0.5 on single crystalline substrates ($SrTiO_3$ and $LaAlO_3$) and also the first deposition by means of rf-sputtering of a cobalt perovskite film.

Our investigation has shown that epitaxial single phase thin films can be successfully deposited by means of rf-magnetron sputtering if the substrates is heated at high temperatures (700°C); lower substrate temperature has shown to lead to multi-phase materials with a low degree of crystallinity.

All the deposited thin films posses a nanocrystalline morphology, even after the post-deposition annealing treatments with average grain size lower than 100 nm. This aspect is of significant interest, particularly when considering the possible applications of these materials as sensors and/or catalysts.

Post-deposition annealing treatments in oxygen are efficient in increasing the oxygen content of the samples, as witnessed by the lattice constant reduction.

All the films deposited on $SrTiO_3$ have shown to be under a tensile strain while those on the $LaAlO_3$ experience a compressive strain thus suggesting that at about 50 nm the films are not fully relaxed even after the thermal treatment. In addition, by reducing the film thickness more strained films have been found for the $x$=0.2 composition. This tuning of the film strain may lead to improved and unexpected magnetic and transport properties as well as induce positive effects on the catalytic/sensing activity of the film. Future work is planned in order to fully characterized the physical properties of the deposited films.

# Acknowledgement

Financial support from the Italian Ministry of Scientific Research (MIUR) by PRIN Projects (2004) is gratefully acknowledged. One of us (L.M.) gratefully acknowledges the financial support of the "Accademia Nazionale dei Lincei". Dr. Oleg Gorbenko is gratefully acknowledged for useful discussion.

"**Supporting Information Available**: Figure showing the two possible epitaxial orientations of the film with respect to the cubic substrate. This material is available free of charge via the Internet at http://pubs.acs.org."

**Figures and Tables Captions**

**Figure 1** – Rietveld refined pattern of $Nd_{0.8}Sr_{0.2}CoO_3$. Red symbols represent the experimental pattern, black line the calculated one while vertical green bars at the bottom of the pattern are the Bragg peaks position. Horizontal blue line shows the difference between the calculated and experimental patterns.

**Figure 2** – X-ray reflectivity spectrum of $NdCoO_3$ deposited for 30' on $SrTiO_3$ (001). The inset highlights the region at low angle.

**Figure 3** – XRD pattern for as-deposited (black line) and annealed (blue line) $Nd_{0.8}Sr_{0.2}CoO_3$ on $SrTiO_3$. Vertical green lines indicate the position of orthorhombic $Nd_{0.8}Sr_{0.2}CoO_3$ peaks, while the red ones refer to the $SrTiO_3$. Inset: enlargement of a small region of the pattern around the main peak.

**Figure 4** – 3D-AFM images for the 50 nm $Nd_{0.8}Sr_{0.2}CoO_3$ thin film after deposition (A) and after the annealing treatment (B).

**Figure 5** – XRD patterns for the $Nd_{1-x}Sr_xCoO_3$ thin films grown on $SrTiO_3$ (001) with $x=0$ (red line), 0.2 (black line) and 0.5 (blue line) around the two main peaks of the patterns.

**Figure 6** – High resolution reciprocal space maps around the asymmetric $(103)^+$ cubic reflection of $SrTiO_3$ for epitaxial $NdCoO_3$ ($x=0$, A), $Nd_{0.8}Sr_{0.2}CoO_3$ ($x=0.2$, B), and $Nd_{0.5}Sr_{0.5}CoO_3$ ($x=0.5$, C). The in-plane lattice parameters and the relaxation degree $R$ are indicated in the Figures.

**Fugure 7** – (a) Rocking curve (omega-scan) around the (002) pseudocubic reflection and (b) reciprocal Space map (RSM) around the (002) cubic reflection of $SrTiO_3$ for the 50 nm thin film of $Nd_{0.5}Sr_{0.5}CoO_3$ showing the low out of plane texture and the absence of truncation rod intensity.

**Figure 8** – XRD pattern for the 50 nm $NdCoO_3$ film on $LaAlO_3$ (001). Inset: comparison between the XRD pattern for the $LaAlO_3$ substrate (red line) and NSCO (black line).

**Figure 9** – 3D-AFM images for the 50 nm $Nd_{0.5}Sr_{0.5}CoO_3$ thin film after deposition (A) and after the annealing treatment (B).

**Figure 10** – XRD patterns for the $Nd_{0.8}Sr_{0.2}CoO_3$ thin films grown on $SrTiO_3$ (001) with thickness 50 nm (blue line), 17 nm (black line) and 8.5 nm (red line) around the two main peaks of the patterns.

**Figure 11** – XRD pattern for NCO deposited at 400°C. Asterisks mark the extra-peaks not directly related to the cobaltite.

**Table 1** – Lattice parameters for the three target materials.

**Table 2** – Out-of-plane parameters of the different films deposited.

# Tables

## Table 1

| *Sample* | *a* | *b* | *c* | *V* |
|---|---|---|---|---|
| $NdCoO_3$ | 5.3369(1) | 7.5530(3) | 5.3484(2) | 215.59(1) |
| $Nd_{0.8}Sr_{0.2}CoO_3$ | 5.3556(3) | 7.5901(4) | 5.3887(2) | 219.05(2) |
| $Nd_{0.5}Sr_{0.5}CoO_3$ | 5.3686(2) | 7.6006(3) | 5.4083(2) | 220.67(2) |

## Table 2

| *Sample* | *Thickness (nm)* | *Substrate* | *Out-of-plane parameter* (Å) |
|---|---|---|---|
| $NdCoO_3$ | 50 | $SrTiO_3$ | 3.766(3) |
| $NdCoO_3$ | 50 | $LaAlO_3$ | 3.780(5) |
| $Nd_{0.8}Sr_{0.2}CoO_3$ | 50 | $SrTiO_3$ | 3.771(3) |
| $Nd_{0.8}Sr_{0.2}CoO_3$ | 50 | $LaAlO_3$ | 3.816(4) |
| Nd$_{0.5}$Sr$_{0.5}$CoO$_3$ | 50 | SrTiO$_3$ | 3.782(3) |
| Nd$_{0.5}$Sr$_{0.5}$CoO$_3$ | 50 | LaAlO$_3$ | 3.820(2) |

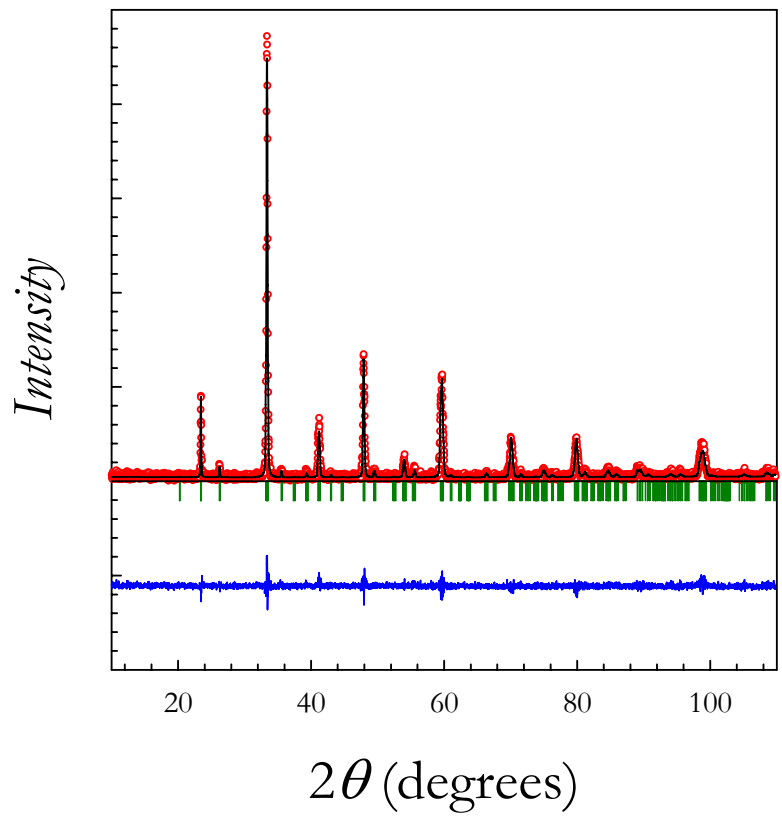

Figure 1

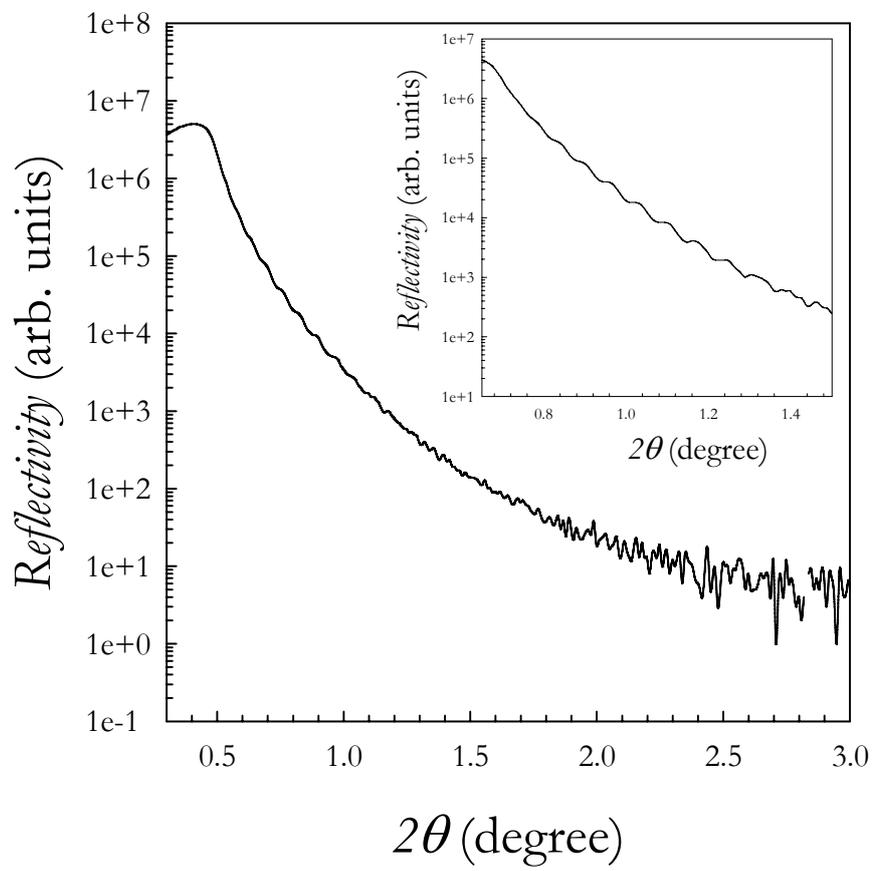

Figure 2

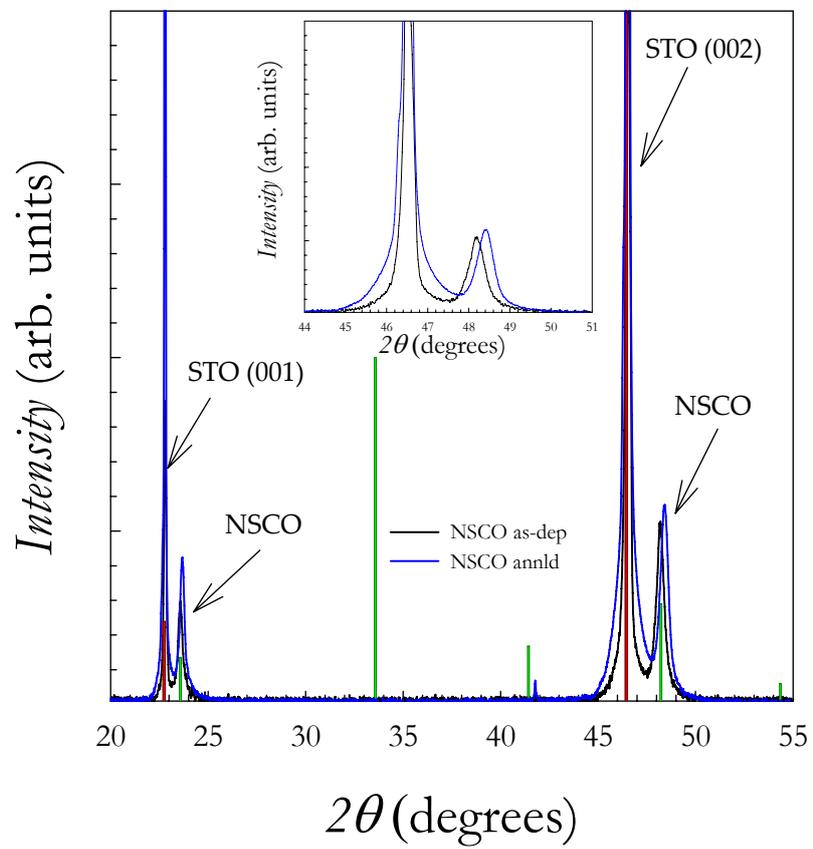

Figure 3

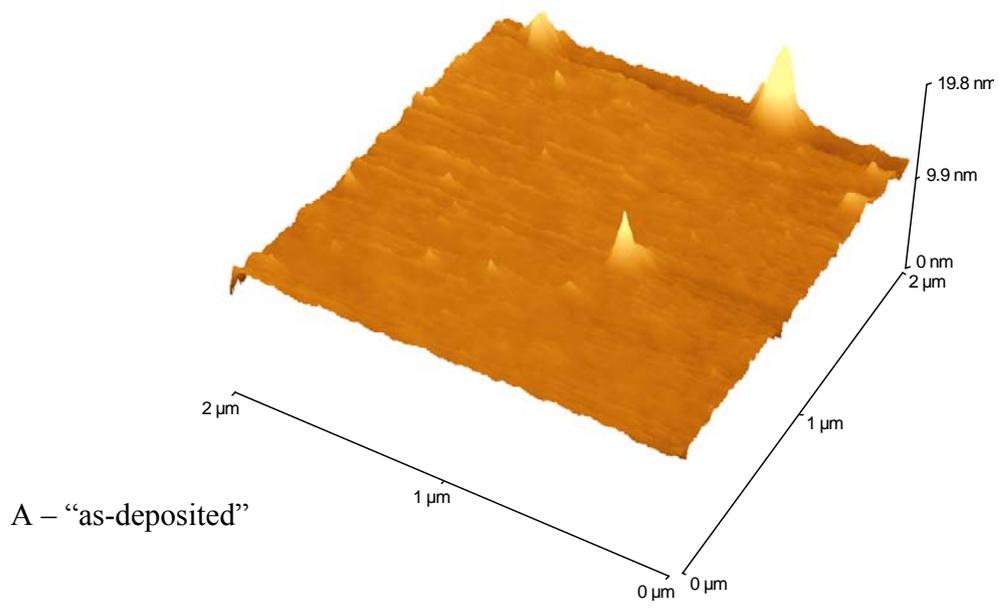

A – "as-deposited"

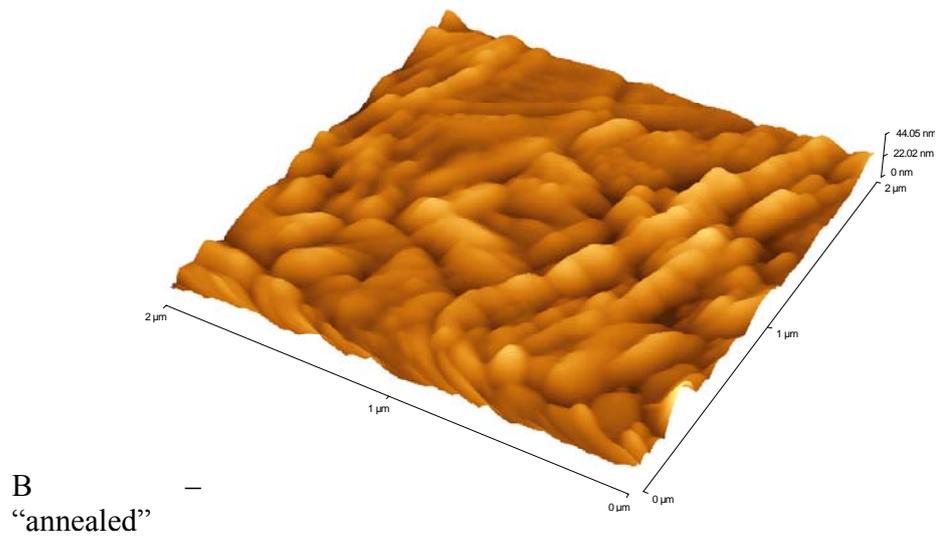

B – "annealed"

Figure 4

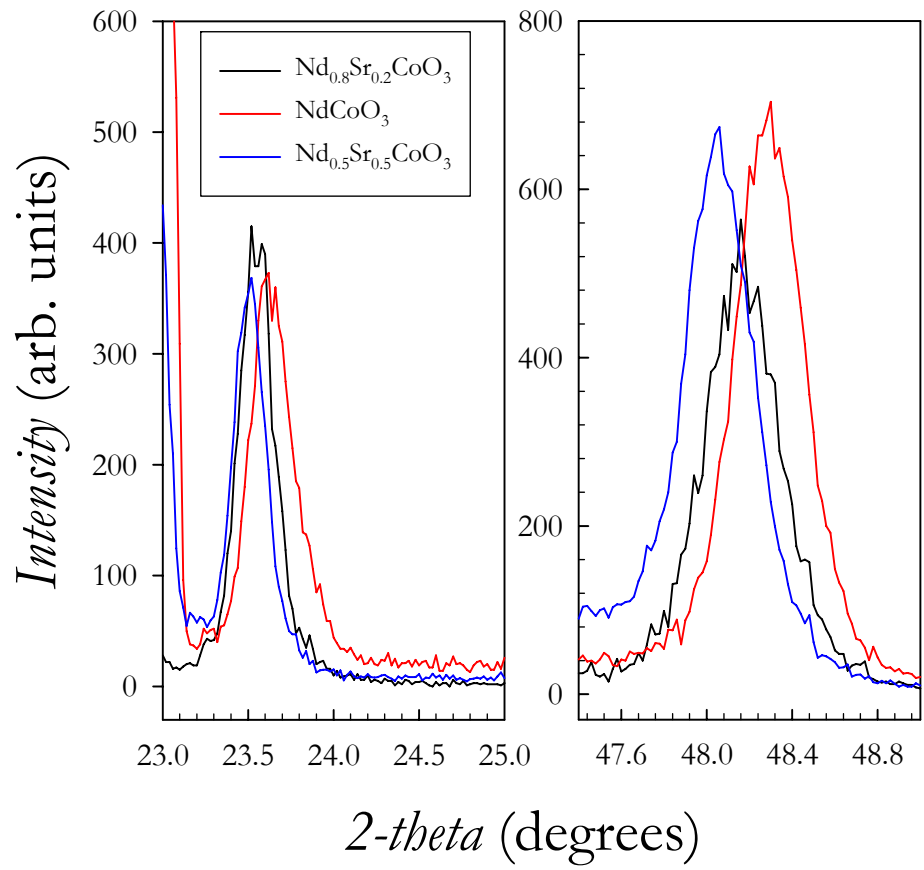

Figure 5

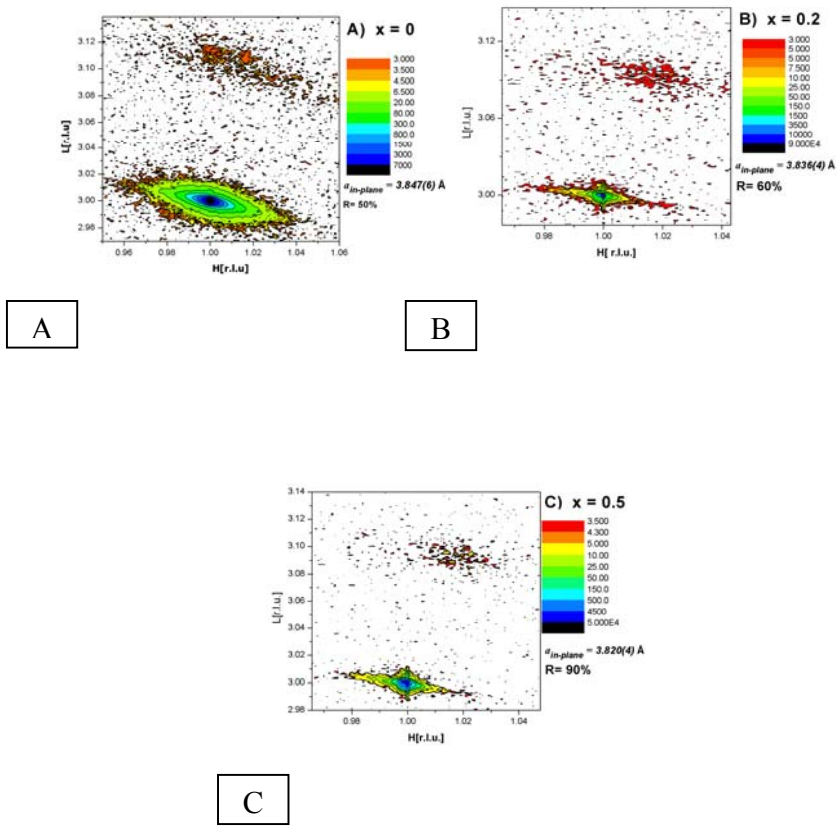

Figure 6

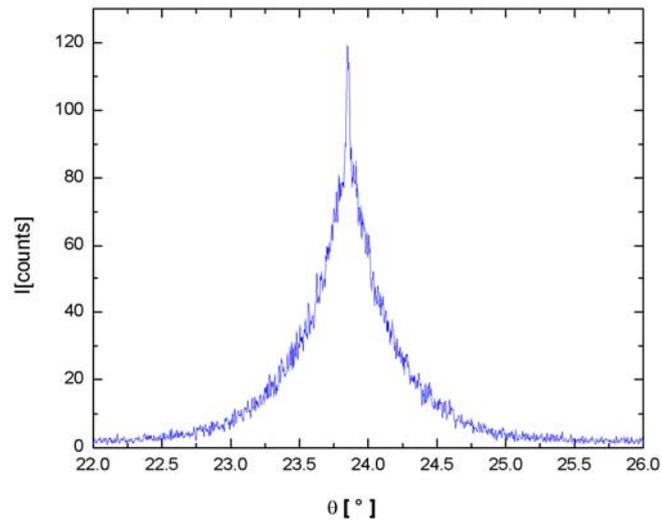

Figure 7a

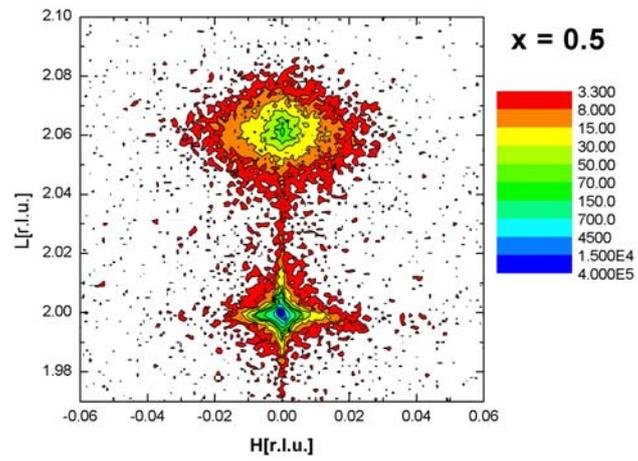

Figure 7b

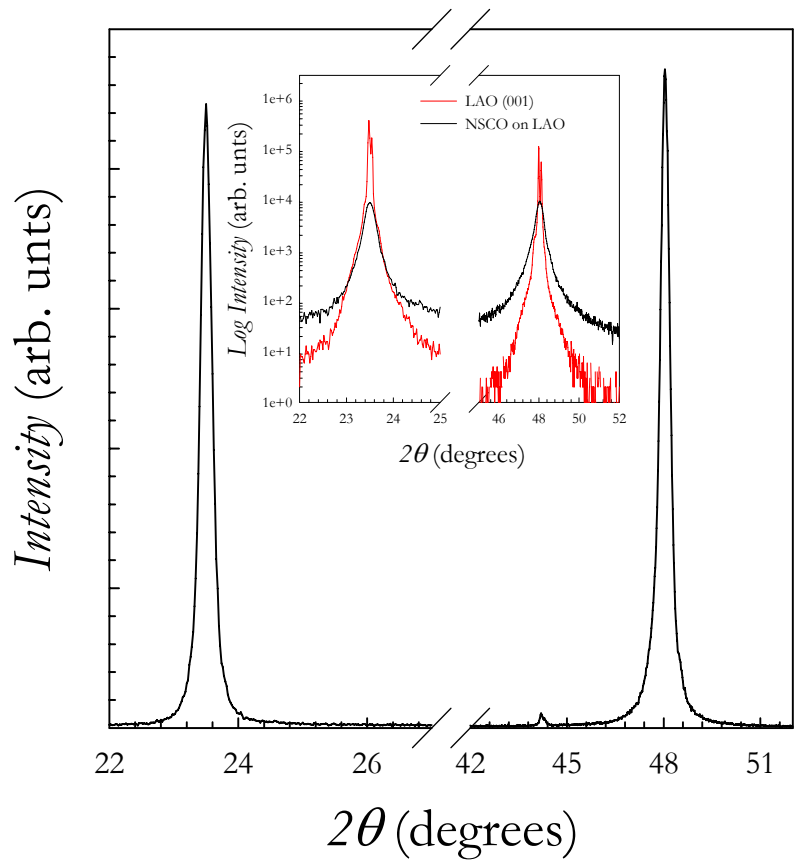

Figure 8

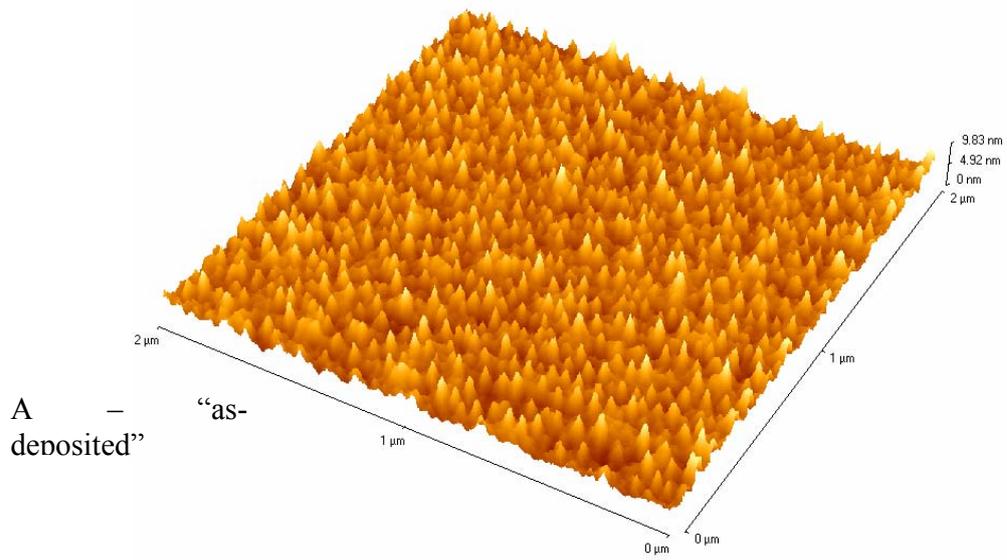

A – "as-deposited"

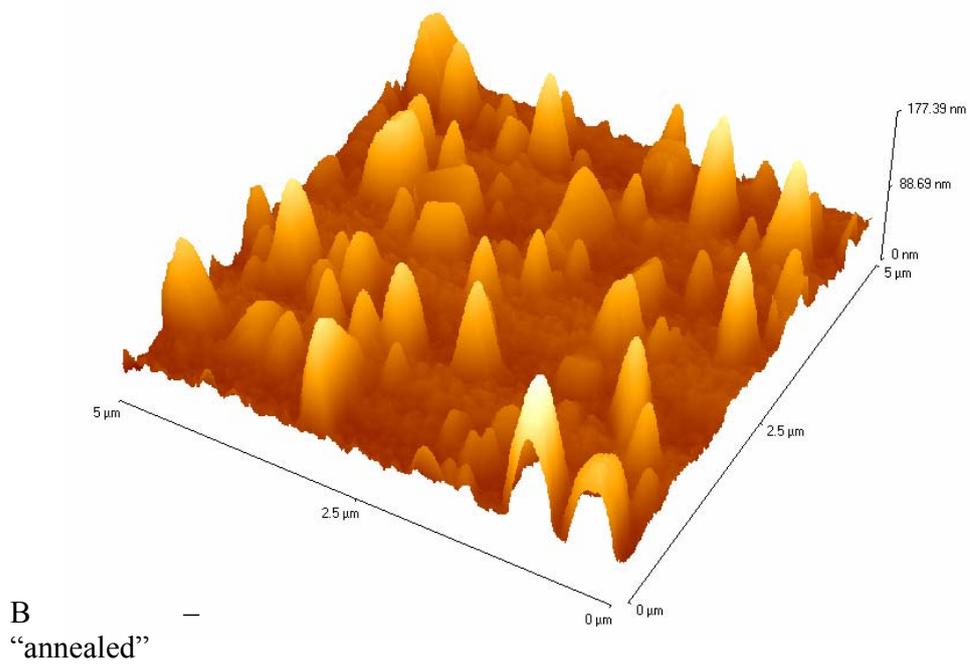

B – "annealed"

Figure 9

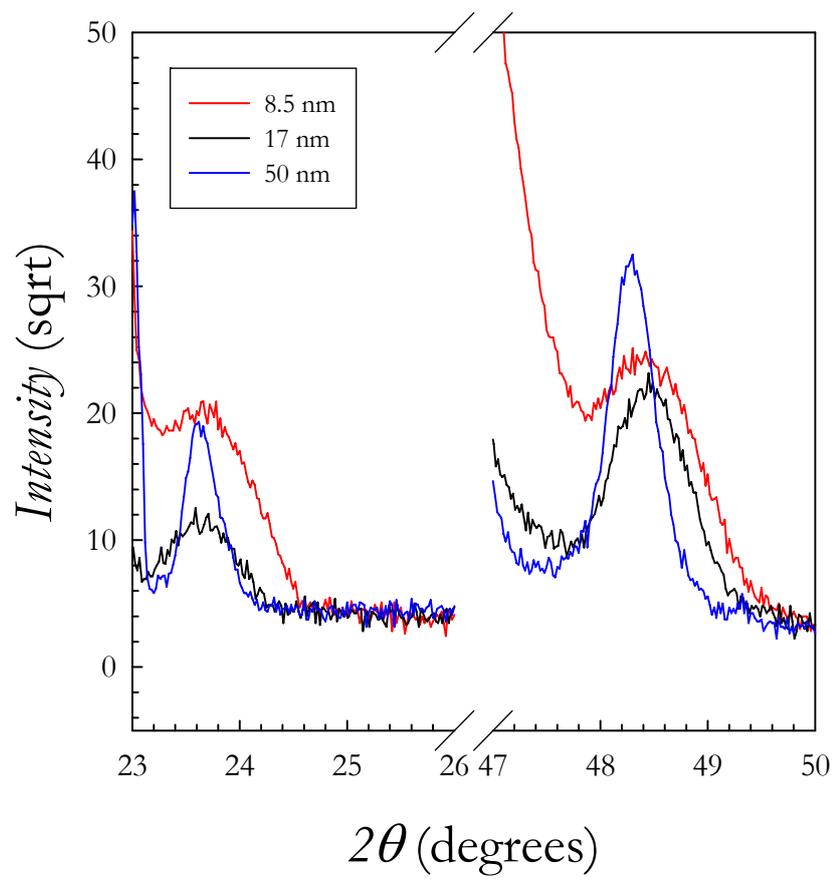

Figure 10

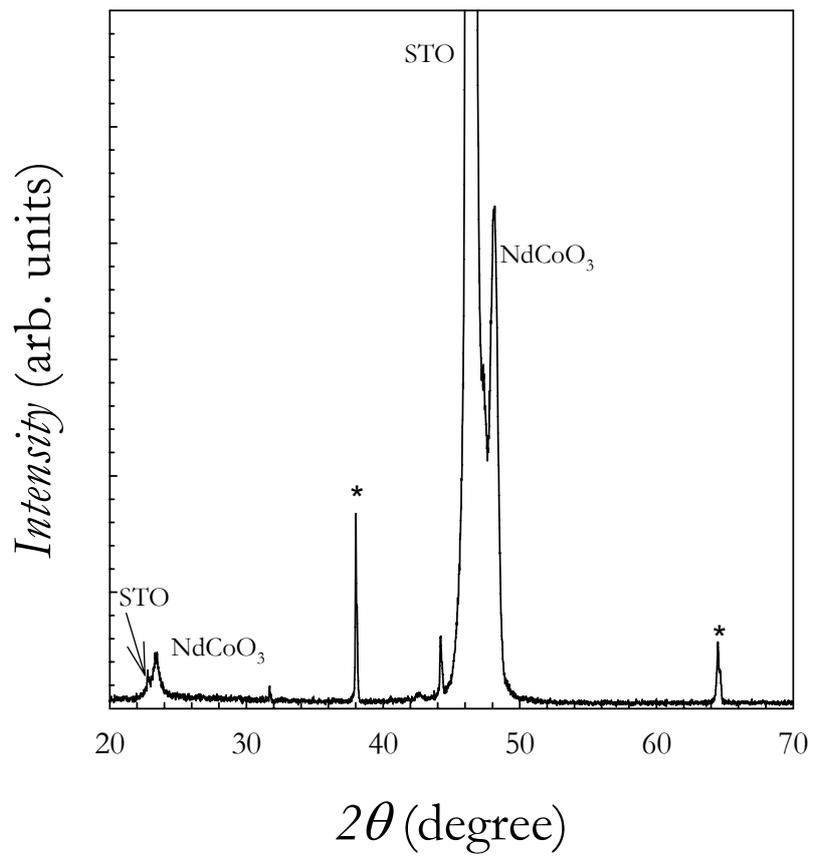

Figure 11